\documentclass[aps,prd,superscriptaddress,twocolumn,nofootinbib,groupedaddress,amsfonts,floatfix]{revtex4}
\usepackage{graphicx,amsmath,amssymb,amstext,amssymb,amsbsy,amsfonts,amsthm,lscape,units}

\newcommand{\pt}{\partial_t}

\newcommand{\lp}{\left(}
\newcommand{\rp}{\right)}

\newcommand{\beq}{\begin{equation}}
\newcommand{\eeq}{\end{equation}}

\newcommand{\mcb}{\mathcal{B}}
\newcommand{\mcl}{\mathcal{L}}

\newcommand{\mplsq}{\, m_{\mathrm{pl}}^2}
\newcommand{\mpl}{\, m_{\mathrm{pl}}}

\newcommand{\gamijl}{e^{4\phi}\bar{\gamma}_{ij}}
\newcommand{\gambar}{\bar{\gamma}_{ij}}

\newcommand{\Aij}{\tilde{A}_{ij}}
\newcommand{\Aiju}{\tilde{A}^{ij}}
\newcommand{\christoff}{\bar{\Gamma}^i}

\usepackage{color}

\begin{document}


\title{Preheating in full general relativity}

\newcommand{\Kenyon}{Department of Physics, Kenyon College, Gambier, Ohio 43022, USA}
\newcommand{\Case}{CERCA/ISO, Department of Physics, Case Western Reserve University, 10900 Euclid Avenue, Cleveland, Ohio 44106, USA}
\newcommand{\Dartmouth}{Department of Physics and Astronomy, Dartmouth College, 6127 Wilder Laboratory, Hanover, New Hampshire 03755, USA}

\author{John~T.~Giblin,~Jr.}
\email{giblinj@kenyon.edu}
\affiliation{\Kenyon}
\affiliation{\Case}
\author{Avery~J.~Tishue}
\email{avery.tishue.gr@dartmouth.edu}
\affiliation{\Kenyon}
\affiliation{\Dartmouth}


\begin{abstract}
We investigate the importance of local gravity during preheating, the nonlinear dynamics that may be responsible for starting the process of reheating the Universe after inflation.  We introduce three numerical methods that study a simple preheating scenario while relaxing gravitational assumptions, culminating in studying the process in full numerical relativity.  We confirm that perturbation theory is no longer valid when one considers modes whose wavelengths are comparable to the size of the horizon at the end of inflation; however, this breakdown does not necessarily lead to a breakdown of the preheating process in nonlinear gravity.  For the specific model we test we find no evidence for the creation of primordial black holes from the instabilities in this model.  Finally, we remark on the opportunity for future numerical study of nonlinear gravitational dynamics in the early Universe.
\end{abstract}

\maketitle


\section{\label{sec1} Introduction}

If inflation occurred in the early Universe, it must have come to an end---at least locally.  While measurements of the cosmic microwave background and large scale structure constrain inflationary models during inflation, the final moments of the inflationary period and the subsequent preheating of the Universe may provide clues for how inflation fits into our models of high energy physics.  The lack of a unique mathematical model of inflation is a compelling reason to search for testable predictions from the inflationary and preheating periods.  Most models of inflation employ a field (or fields) whose homogeneous value(s) determines the final dynamics of this epoch; in many of these, the field ends up oscillating about a minimum of the potential.  Frequently in these scenarios, nonlinear processes take over to accelerate the decay of the inflaton field in a family of scenarios known as {\sl preheating}, see, e.g., Ref. \cite{Amin:2014eta} for a review.

While the processes of the field sectors of preheating have been examined closely throughout the past decades, the effects of {\sl gravity} during preheating have not yet been extensively studied.  At the same time, there exist some exciting possibilities from the preheating period.  Large, nonthermal, and nonlinear inhomogeneities are characteristic of this period and detailed numerical study is needed to understand the role of this physics. 

In this paper we aim to do a first investigation of nonlinear gravity during the violent process of preheating in the early Universe. 

To date, cosmological perturbation theory (CPT) has been an effective and predictive way to study departures from a purely homogeneous and isotropic, Friedmann-LeMa\^itre-Robertson-Walker (FLRW), universe.  All previous work studying gravity during and after preheating exists in this scenario. The first was done using {\sc DEFROST}, \cite{Frolov:2008hy}, where it was shown that local gravitational effects seem to be small, an observation that has been confirmed in other numerical simulations.  More recent work continues to study the generation of linear gravitational perturbations during preheating, \cite{Huang:2011gf,Lozanov:2019ylm}, where gravitational effects show promise in helping to understand the rich phenomenology of this period.  

However, CPT is inherently limited, by construction, to a linearized treatment of gravity and is not able to resolve the strong-field regime. Equally important is that linearized gravity, by definition, does not allow for gravitational modes to couple, hiding potentially important aspects of the preheating process.  As such, the search for new physics beyond the perturbative regime strongly motivates the application of full general relativity (GR) to the study of cosmology.  Applying nonlinear gravity to cosmological scenarios is a recent advancement in high resolution and accurate numerical simulations.  The formalisms employed to do full numerical relativity \cite{Pretorius:2005gq,Campanelli:2005dd,Baker:2005vv,Garfinkle:2003bb,Baumgarte:1998te,Shibata:1995we} have been recently applied to late universe scenarios \cite{Giblin:2015vwq,Bentivegna:2015flc,Giblin:2016mjp,Bentivegna:2016stg,Macpherson:2016ict,Giblin:2017juu,East:2017qmk,Wang:2018qfr,Macpherson:2018akp,Macpherson:2018btl,Giblin:2018ndw} as well as preinflationary scenarios \cite{East:2015ggf,East:2016anr,Clough:2017efm,Bloomfield:2019rbs} and oscillons \cite{Muia:2019coe}.  
Other work along these lines include adding gravitational effects for scalar metric perturbations, beyond linear order, \cite{BasteroGil:2007mm,BasteroGil:2010nm} during the preheating period.  Other works include fully nonlinear gravity, either in one dimension \cite{Easther:1999ws} or in a restricted gauge in Ref. \cite{BasteroGil:2011zza}. 

In CPT, the linearized Einstein's equations couple spatial derivatives of the Newtonian gravitational potential $\Phi$ to local inhomogeneities in the energy density, $\delta\rho$.  When inhomogeneities grow, the gravitational response may be large.  

The idea that preheating can amplify gravitational degrees of freedom dates back to early work \cite{Nambu:1996gf,Taruya:1997iv} where the authors noticed that gravitational modes can be excited via parametric instabilities.  This work was significantly extended by Refs. \cite{Bassett:1998wg,Bassett:1999mt,Bassett:1999ta,Tsujikawa:2002nf}, where the authors set out to understand where linear perturbation theory breaks down and point out that nonlinear gravity is necessary to study the gravitational effects of preheating \cite{Bassett:1999mt}. 

More recently, it has been noted that overdensities on fixed subhorizon scales might collapse due to nonlinear gravitational effects \cite{Jedamzik:2010dq,Jedamzik:2010hq}---a generic feature of preheating which could lead to the formation of primordial black holes, e.g., Ref. \cite{Martin:2019nuw}.  Additionally, it was shown in Ref. \cite{Easther:2010mr} that an instability exists for modes near the Hubble scale at the end of inflation.  These modes provide the dominant contribution to the density contrast $\delta(t,\vec{x})= \delta \rho(t,\vec{x})/\left<\rho\right>$, causing it to become large.  This may signal the breakdown of linear perturbation theory and the need for a nonlinear treatment.  

Our primary goal here is to increase the robustness of gravitational approximations during the preheating stage, culminating in studying the problem in full numerical relativity.  We begin with a standard implementation of the Grid and Bubble Evolver {\sc GABE} in an FLRW universe before including localized linear gravity in Newtonian gauge. We then introduce {\sc GABERel}, an adaptation of {\sc GABE} \cite{Child:2013ria} that numerically evolves the full set of Einstein's field equations on an expanding background. We utilize {\sc GABERel} to evolve nonlinear inflaton modes in the postinflationary Universe during parametric resonance preheating. This provides a precise numerical treatment of the nonlinear effects that lead to the breakdown of coherent oscillations in the postinflationary Universe and demonstrates the ability of {\sc GABERel} to move beyond limitations of linear perturbation theory.   We confirm prior work that predicts the breakdown of perturbation theory at the end of infation; however, we go further to show that full numerical relativity safely controls these inhomogeneities which do not end up being catastrophic: by either causing the coherent modes of the inflaton to break down or by seeding primordial black holes. 

We also present the formalism one can use to calculate the (first-order) gauge-invariant Bardeen potentials in the framework of full GR. We use this as a test of {\sc GABERel}'s robustness and to compare these potentials to the Newtonian simulation in an equivalent preheating context. We also use GABERel to evolve a single black hole of mass $M$ to assess its ability to resolve strong-field dynamics to demonstrate the accuracy of this new technique.

This paper is organized as follows. In Sec.~\ref{model} we introduce our model of preheating and its numerical implementation in three simulations with increasingly comprehensive treatments of gravitational dynamics. In Sec.~\ref{numericalsims} we outline our initialization procedure and give computational details. We present our results in Sec.~\ref{Results} and a discussion in Sec.~\ref{discussion}. Appendix~\ref{bardeenfromBSSN} contains information regarding how to calculate the Bardeen potentials in a fully general-relativistic simulation. Appendix~\ref{verify} outlines methods by which we validate the accuracy of {\sc GABERel}'s gravitational dynamics. 

\section{\label{model} Model}

We are primarily interested in the effects of gravity on preheating; hence, we study a canonical model of preheating with two scalar fields:  the inflaton, $\varphi_1$, and a coupled, massless, matter field, $\varphi_2$.  For convenience we use upper-case latin letters, $I$, $J$, etc, as a field index to simplify notation when we need to sum over fields; we use $\varphi_I$ when we need to generally refer to one or all of the scalar fields and repeated field indices imply a sum.  The matter Lagrangian for our system is 
\beq
\mcl _{\mathrm{m}} =  \left(\frac{1}{2}\partial^\mu \varphi_I \partial_\mu \varphi_I\right)  - V(\varphi_I)
\label{lagrangian}
\eeq
with a potential  
\beq
V(\varphi_I) = \frac{1}{2}m^2 \varphi_1^2  +\frac{1}{2}g^2 \varphi_1^2 \varphi_2^2.
\label{preheatmodel}
\eeq
Throughout this work, we take the coupling to be $g^2 = 2.5 \times 10^{-7}$ and $m = 10^{-6}\, \mpl$ as a toy model that has been studied extensively in the literature \cite{Traschen:1990sw}.  This is an interesting model both because it has been repeatedly used as a benchmark preheating scenario and is widely recognized, but also because it exhibits {\sl broad resonance}---see, e.g., Fig.~3 in Ref.  \cite{Amin:2014eta} where linear analysis predicts efficient preheating. Since the dynamics of this model are so well known, it is an ideal case to study as a first try even though this inflationary model is disfavored \cite{Aghanim:2018eyx} and this specific model of preheating needs extensions to completely deplete the energy in the inflaton, e.g., Ref.  \cite{Dufaux:2006ee}. In this model, when inflation ends the vast majority of the energy density of the Universe is trapped in the inflaton condensate.  As the homogeneous mode of the inflaton field oscillates at the minimum of the potential it parametrically amplifies modes of the matter field until the $\varphi_2$ particles scatter into the $\varphi_1$ field and nonlinear physics leads to the breakdown of the condensate. We want to be careful here to delineate the different phases of the process.  The first of these phases, I, is where the $\varphi_2$ field is being amplified, but the inflaton condensate is mostly unaffected by any backscattering.  The next phase, II, is characterized by backscattering onto $\varphi_1$ and an amplification of the variances of both the $\varphi_1$ and $\varphi_2$ fields.  The final phase, III, sees dramatic nonlinear amplifications of both fields; this phase is important as significant power is transferred between different modes.  When studying this process on a finite grid, it is important to note that we almost always reach a stage at which significant power is transferred to the smallest-resolvable scales.  We will note this moment in the following sections as it becomes more dangerous as the dynamics become more complicated. 

Since we are interested in gravitational effects, and want to be able to include any effects that arise when $\ddot{a} = 0$, we start our simulations one $e$-folding before inflation ends.  In this model, that corresponds to a (homogeneous) field value of $\phi_0 \approx 0.415\, m_{\rm pl}$ and velocity $\dot{\phi}_0 \approx -0.154\,\mpl$.  Inflation ends when the field passes $\phi_0 \approx 0.201\, \mpl$. Throughout the work here we denote the subscript ``0" to refer to quantities evaluated at the beginning of the simulation, i.e. we take $a_0 = e^{-1}$, and the subscript ``$*$" will refer to quantities evaluated at the end of inflation, $a_* = 1$.

In all simulations, we will initialize our scalar field modes to be in the Bunch-Davies vacuum, 
\beq
\left<\left|\varphi_{I}(k)\right|^2\right> = \frac{1}{2\omega}
\label{bunchdavies}
\eeq
where $\omega = \sqrt{k^2 + m_{\rm eff}^2}$. In an expanding universe, this approximation is only good for modes (well) inside the horizon---however there are known adaptations that work for horizon-sized modes \cite{Frolov:2008hy}.  Since we start our simulations before the end of inflation, it allows the entire (or most of the) box to be subhorizon on the initial slice. In doing so we can trust our initial conditions while still allowing the fields to evolve to be superhorizon by the time inflation ends---knowing that there could be instabilities on the scale of $k/a \sim H_*$.  

In the next three subsections we present three increasingly comprehensive treatments of the gravitational physics in this model.  First we introduce the canonical treatment in a rigidly expanding spacetime, the FLRW limit.  We then study the problem in perturbation theory in Newtonian gauge in CPT, followed by an introduction of the fully nonlinear methods using the BSSN formalism \cite{Baumgarte:1998te,Shibata:1995we}.

\subsection{The FLRW limit}
\label{FLRWlimit}

\begin{figure}[ht]
\centering
\includegraphics[width=\columnwidth]{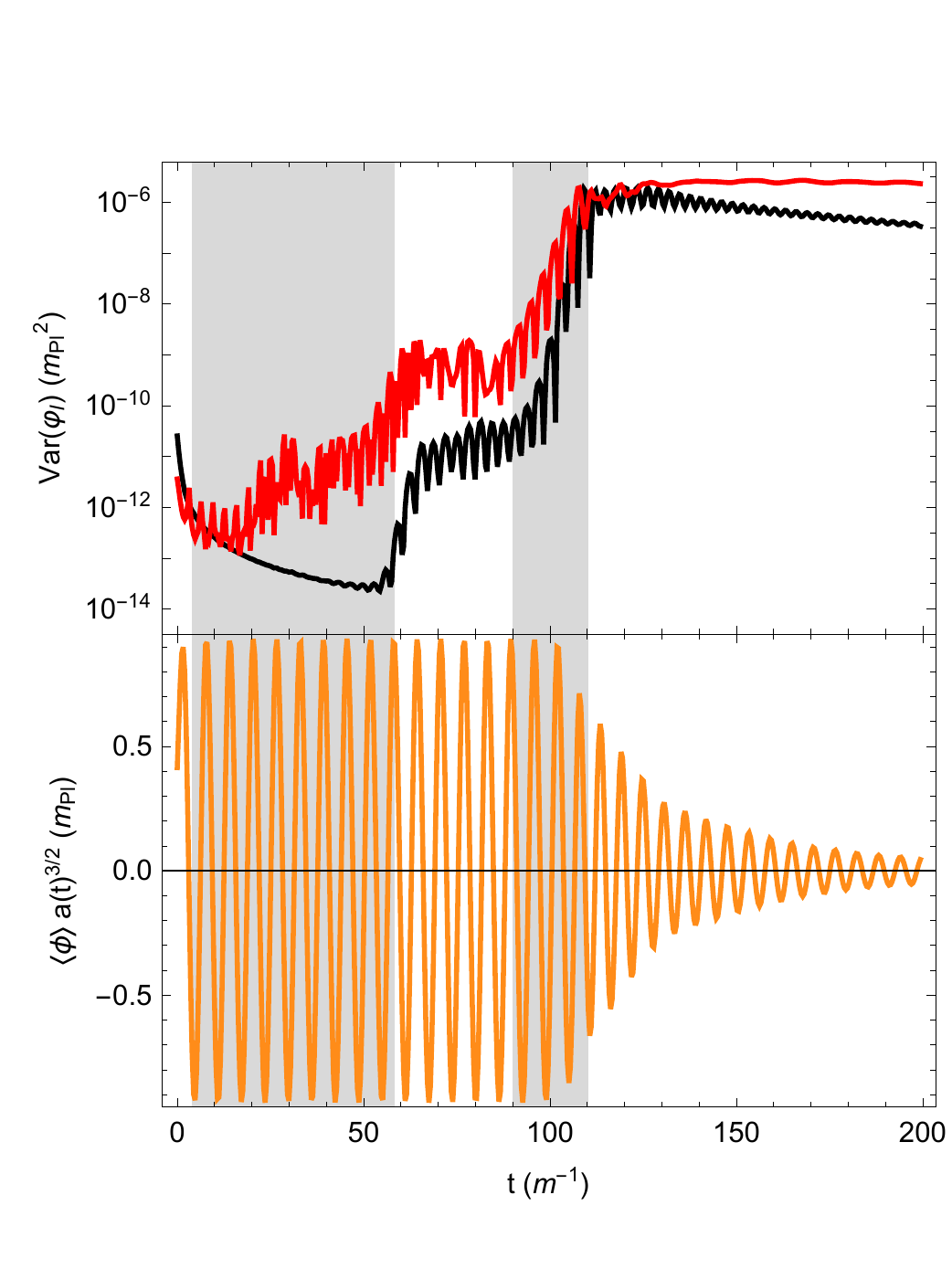} 
\caption{The stages of parametric resonance preheating in the FLRW limit. The top panel illustrates the evolution of the variances of the scalar fields $\varphi_1$ (black) and $\varphi_2$ (red). In the first shaded region, inflaton oscillations amplify modes of the $\varphi_2$ field through the potential coupling, but backscattering has yet to break down the inflaton condensate. In between the two shaded regions, there are enough $\varphi_2$ particles to backscatter into the inflaton field, causing the variances of both fields grow. In the second shaded region, both field variances grow rapidly indicating power is being distributed over many modes. These three phases of field variances are characteristic of parametric resonance preheating. In the bottom panel, the orange curve tracks the initially matterlike evolution of the inflaton condensate as it oscillates around the potential minimum.}
\label{prehstages}
\end{figure}

In general, studies of preheating are done in the FLRW regime, where the line element is
\beq
\label{FLRWmet}
g_{\mu\nu}^{\rm FLRW} = {\rm diag} (-1, a^2(t), a^2(t), a^2(t))
\eeq
and the dynamics of the scale factor are determined by the 00-component of Einstein's equations, Friedmann's equation,
\beq
H^2 \equiv \left(\frac{\dot{a}}{a}\right)^2 = \frac{8\pi}{3 m_{\rm pl}^2} \left<\rho\right>
\eeq
where the $\left<\ldots \right>$ denote an average over a constant-time hypersurface, and 
\beq
\rho \equiv - T_0^0 = V(\varphi_I) + \sum_I \lp \frac{1}{2} \dot{\varphi_I}^2 + \frac{\left(\nabla \varphi_I\right)^2}{2a^2}\rp.
\eeq
As usual each of the scalar fields evolve according to the Klein Gordon equation,
\beq
\label{KG}
\ddot{\varphi_I} + 3 H \dot{\varphi_I} - \frac{\nabla^2 \varphi_I}{a^2} +\frac{\partial V}{\partial \varphi_I} = 0.
\eeq
To be comprehensive, Eq.~(\ref{KG}) implies similar---but crucially different---equations of motion for the inflaton and matter field in this preheating model,
\begin{align}
\ddot{\varphi_1} &+ 3 H \dot{\varphi_1} - \frac{\nabla^2 \varphi_1}{a^2} + g^2 \varphi_1 \varphi_2^2 +m^2 \varphi_1  = 0 \\
\ddot{\varphi_2} &+ 3 H \dot{\varphi_2} - \frac{\nabla^2 \varphi_2}{a^2} + g^2 \varphi_1^2 \varphi_2 =0.
\end{align}
Figure \ref{prehstages} illustrates the importance of these differences in this canonical process for a simulation with $L_* = 5 \, m^{-1}$.  While this box is slightly superhorizon at the end of inflation it becomes subhorizon very quickly after;  shaded regions in this figure show the three phases of preheating as described above.  To track the production of particles across the box, we use the variance of the fields,
\beq
{\rm Var(\varphi_I)} = \sqrt{ \langle \varphi_I^2\rangle - \langle \varphi_I \rangle^2 },
\eeq
as a way to assess inhomogeneity as a function of time. In the FLRW regime, Friedmann's equation contains all of the gravitational physics and there is no local gravitational collapse; the fields simply evolve in an expanding background, preheating the Universe.

\subsection{Cosmological perturbation theory}
\label{CPT}

To include the effects of local gravity, it is natural to next look to cosmological perturbation theory.  In this limit we introduce a scalar perturbation to Eq.~(\ref{FLRWmet}), in our case in Newtonian gauge \cite{Bardeen:1980kt,Ma:1995ey,Weinberg:2008zzc},
\beq
\label{CPTmet}
g^{\rm CPT}_{\mu\nu} = \left(\begin{array}{cc} -(1+2\Phi) & 0 \\ 0 &  a^2(t)(1-2 \Psi)\end{array}\right)
\eeq
The metric perturbations $\Phi$ and $\Psi$ are the Newtonian gravitational potentials. They are equal when the scalar anisotropic stress vanishes, an assumption we will make. While this is not strictly true for a scalar field source, it is a good approximation when the stress-energy tensor is almost diagonal.  We also note that we are not choosing to use the (first-order) gauge invariant Mukhanov-Sasaki variables \cite{Mukhanov:1990me}, as is commonly done in CPT; here we wish to keep the metric and field perturbations separate with an eye toward going beyond the linear analysis. 
The energy density can be decomposed at linear order into a background value and a perturbation,
\begin{equation}
    \rho =  \left< \rho \right> + \delta \rho,
\end{equation}
so that we can study Einstein's equations order by order. The 00 component yields Friedmann's equation,  Eq.~(\ref{FLRWmet}), at zeroth order,  and 
\begin{equation}
\label{poisson1}
    \nabla^2 \Phi - 3H a^2 \lp \dot{\Phi} + H \Phi \rp= 4\pi G a^2 \delta \rho 
\end{equation}
at first order.  It is common to drop the second term of Eq.~(\ref{poisson1}) in the nonrelativistic limit; however, since our sources are scalar fields, we can calculate the second term by simultaneously solving the divergence of the $0i$ terms of Einstein's equations, $\delta^{ij}\partial_i G_{0i} = 8\pi/m^2_{\rm pl} \, \delta^{ij}\partial_i T_{0i}$,
\beq
\label{poisson2}
\nabla^2 \left(\dot{\Phi} + H \Phi \right) = \sum_I \frac{4\pi}{\mpl^2}  \delta^{ij} \partial_i \left(\partial_j \varphi_I \partial_0 \varphi_I \right).
\eeq
The effects of local gravity enters into the equations of motion for the scalar fields, $\partial_\mu \left(\sqrt{-g} g^{\mu\nu}\partial_\nu \varphi_I \right) = \sqrt{-g}\, dV/d \varphi_I$, resulting in 
\begin{align}
 \ddot{\varphi}_I =  &  \lp \frac{4\Phi +1}{a^2}  \rp \nabla^2 \varphi_I \nonumber  \\ 
& +(4\dot{\Phi} -3 H)\dot{\varphi}_I - (2 \Phi +1)\frac{\partial V}{\partial \varphi_I}
\end {align}
In practice we use a spectral method to solve Eqs.~(\ref{poisson1}) and (\ref{poisson2}) simultaneously. We take the Fourier transform of the right-hand side of these two equations and then invert the Laplacians in momentum space to get the Fourier transform of $\Phi$ as well as the combination $(\dot{\Phi} + H\Phi)$; we then inverse Fourier transform these to get $\Phi$ and $\dot{\Phi}$ in configuration space.  
 
Local gravity during preheating has been studied in Refs. \cite{Frolov:2008hy,BasteroGil:2007mm,BasteroGil:2010nm}, where the size of the Newtonian potential was found to be small. On the other hand, an examination of Eq.~(\ref{poisson1}) shows that the magnitude of the Newtonian potential depends on the volume of the simulation.  As the wavelength of a density perturbation approaches to the Hubble scale $k \rightarrow H$, the resulting size of the gravitational response $\Phi$  grows, even if $\delta \rho/\langle \rho \rangle$ is constant.  Therefore allowing density perturbations to exist at near-horizon scales (by including more modes in the simulation) amplifies the size of the Newtonian potential. 

This is the case in Ref. \cite{Easther:2010mr}, where there is a mechanism through which the Universe becomes inhomogeneous at the Hubble scale at the end of preheating. In this scenario inflaton modes with $k \sim a_* H_*$ contribute significantly to the extrinsic curvature as well as the density contrast, eventually spoiling linearity. We demonstrate this effect in Sec.~\ref{Results}, Fig.~\ref{boxsizev2}  by including these modes in the CPT simulation; after enough time in the phase of coherent oscillations, roughly $\sim 2$-$3$ $e$-foldings, these modes grow nonlinearly and the validity of the perturbative treatment breaks down. This breakdown, discussed at length in Sec.~\ref{Results}, necessitates a fully general-relativistic treatment of the postinflationary Universe's gravitational dynamics, raising the possibility of primordial gravitational wave production during preheating as well as the nonlinear growth of density perturbations in a strong gravitational regime \cite{Martin:2019nuw}.

\subsection{Fully nonlinear gravity}

The BSSN Formalism---an adaptation of the ADM formalism of GR \cite{Arnowitt:1959ah,Baumgarte:1998te,Shibata:1995we}---utilizes a 3+1 decomposition of the field equations, with the metric parametrized as
\beq
\label{BSSNmet}
g^{\rm BSSN}_{\mu\nu} = \left(\begin{array}{cc} -\alpha^2 + \beta_l \beta^l & \beta_j \\ \beta_i & e^{4\phi}\bar{\gamma}_{ij}\end{array}\right).
\eeq
The variables $\alpha$ and $\beta^i$ are called the lapse and the shift and parametrize gauge degrees of freedom. The spatial metric $\gamma_{ij}$ is conformally rescaled, $\gamma_{ij} = \gamijl$, such that the conformal metric has unit determinant $\mathrm{det}\left| \bar{\gamma}_{ij} \right| = 1$. This notation will be general so that quantities written with overbars are related to the conformal spatial metric, $\bar{\gamma}_{ij}$. Then, the field equations can be expressed as a system of first order differential equations in these metric variables,
\begin{align}
\partial_t \phi = & -\frac{1}{6} \alpha K +\beta^i \partial_i \phi + \frac{1}{6}\partial_i \beta^i   \\
\partial_t \gambar = &-2 \alpha \Aij + \beta^k \partial_k \gambar + \bar{\gamma}_{ik} \partial_j \beta^k  \nonumber \\
& + \bar{\gamma}_{kj} \partial_i \beta^k - \frac{2}{3}\bar{\gamma}_{ij} \partial_K \beta^k  \\
\partial_t K =& \gamma^{ij} D_jD_i \alpha + \alpha \lp \Aij \Aiju +\frac{1}{3}K^2 \rp \nonumber  \\ 
& + 4\pi \alpha \lp \rho +S \rp +\beta^i \partial_i K  \\
\partial_t \Aij  =& e^{-4\phi} \lp -  D_jD_i \alpha + \alpha ( R_{ij}
- 8\pi S_{ij}) \rp^{TF} \nonumber  \\
&  + \alpha \lp K \Aij -2 \tilde{A}_{il} \tilde{A}^l_j \rp + \beta^k \partial_k \Aij \nonumber \\
& +\tilde{A}_{ik}\partial_j \beta^k + \tilde{A}_{kj}\partial_i \beta^k -\frac{2}{3}\Aij \partial_k \beta^k 
\end{align}
In addition to these evolution equations, the BSSN formalism improves numerical stability by defining an auxiliary variable $\bar{\Gamma}^i \equiv \bar{\gamma}^{jk} \bar{\Gamma}^i_{jk} = -\partial_j \bar{\gamma}^{ij}$, which is evolved independently, 
\begin{align}
\label{christoffevol}
\partial_t \bar{\Gamma}^i = & -2\Aiju \partial_j \alpha +2 \alpha ( \bar{\Gamma}^i_{jk} \tilde{A}^{kj} - \frac{2}{3}\bar{\gamma}^{ij} \partial_j K \nonumber \\
&  - 8\pi \bar{\gamma}^{ij} S_j + 6\Aiju \partial_j \phi ) + \beta^j \partial_j  \bar{\Gamma}^i   \nonumber \\
& -  \bar{\Gamma}^j \partial_j \beta^i +\frac{2}{3}  \bar{\Gamma}^i  \partial_j \beta^j +\frac{1}{3}\bar{\gamma}^{li} \partial_l \partial_j \beta^j \nonumber \\
&  + \bar{\gamma} ^{lj} \partial_j \partial_l \beta^i 
\end{align}
Using the auxiliary variables effectively promotes the ``divergence" of the conformal metric to an independent variable so that its second spatial derivatives do not need to be directly calculated via lattice stencils.  This promotion makes the problem hyperbolic which, hopefully, causes numerical noise to remain bounded in the simulation. We see this same technique in a moment when we discuss the scalar field evolution. 

The $0\mu$ components of the field equations furnish four constraint equations which must be satisfied at all times throughout the system's evolution to ensure that the numerical simulation remains a valid solution.  These read
\begin{align}
\label{Hconstr}
\mathcal{H} \equiv 0 = & \bar{\gamma}^{ij} \bar{D}_i \bar{D}_j e^{\phi} - \frac{e^{\phi}}{8}\bar{R} +\frac{e^{5\phi}}{8}\Aiju \Aij \nonumber \\
& - \frac{e^{5\phi}}{12}K^2 + 2\pi e^{5\phi} \rho, \\
\label{Mconstr}
\mathcal{M}^i \equiv 0 = & \bar{D}_j \lp e^{6\phi} \Aiju  \rp -\frac{2}{3}e^{6\phi} \bar{D}^i K - 8\pi e^{10 \phi} S^i,
\end{align}
in which 
\begin{equation}
    S^i = - \gamma^{ij}n^{a}T_{aj},
\end{equation}
where the vector $n^a=(\alpha^{-1},-\alpha^{-1}\beta^i)$ is normal to the spatial hypersurface.

Satisfying these {\sl Hamiltonian}, Eq.~(\ref{Hconstr}), and {\sl momentum} Eq.~(\ref{Mconstr}), constraints---or showing that they are bounded---is a central challenge of performing fully relativistic simulations, which we discuss in Appendix~\ref{verify}.

Because the lapse and shift are gauge variables, we are free to define how they evolve; this process fixes a {\sl slicing} of spacetime which makes it possible to choose the coordinates best fit for different problems. One common choice, {\sl geodesic slicing} or {\sl synchronous gauge}, fixes $\alpha =1$ and $\beta^i = 0 $, so an observer at fixed spatial coordinates travels along a geodesic. Another extensively studied choice for the evolution of $\alpha$ is a family of coordinate systems called {\sl $1+\log$ slicing} 
\beq
\label{1pluslognormal}
\lp \pt -\beta^j \partial_j \rp \alpha = -2 \alpha  K.
\eeq
Here we almost exclusively employ a variant of Eq.~(\ref{1pluslognormal}), where the extrinsic curvature is replaced by its deviation from its average on our spatial hypersurfaces $\langle K \rangle$,
\beq
\label{1pluslogcond}
\pt \alpha = -2 \alpha \lp K - \langle K \rangle \rp.
\eeq
When using Eq.~(\ref{1pluslogcond}), we will use the {\sl hyperbolic Gamma-driver} condition for the shift,
\begin{align}
\label{hypergamdriv1}
\pt \beta^i  &= \frac{3}{4}B^i  \\
\label{hypergamdriv2}
\pt B^i  &= \pt \bar{\Gamma}^i - \eta B^i .
\end{align}
The combination of Eqs.~(\ref{1pluslognormal}), (\ref{hypergamdriv1}), and (\ref{hypergamdriv2}) have proven to be highly robust choices suitable for black hole simulations.  The parameter $\eta$ is a constant often related to the total mass in the simulation \cite{Baumgarte:2010ndz}. In this work we find good numerical results with $\eta \sim 50 \, m$.

Following previous work \cite{Baumgarte:2010ndz,Clough:2015sqa}, the scalar field equations are split into two first-order differential equations where the momentum of the fields are replaced with $\Pi_I$,
\begin{align}
\Pi_I \equiv &\frac{1}{\alpha}\lp \pt \varphi_I - \beta^i \partial_i \varphi_I \rp \\
\pt \Pi_I =&  \beta^i \partial_i \Pi_I +\gamma^{ij} \lp \partial_i \partial_j \varphi_I + \partial_j \varphi_I \partial_i \alpha \rp \nonumber  \\
&  +\alpha \lp K\Pi_I - \gamma^{ij} \Gamma^k_{ij} \partial_k \varphi_I -\frac{\partial V}{\partial \varphi_I} \rp.
\end{align}
As mentioned earlier when discussing $\bar{\Gamma}^i$ in Eq.~(\ref{christoffevol}), we also promote the spatial derivative of the scalar field to an independent variable, $\psi_{iI} \equiv \partial_i \varphi_I$, which evolves according to 
\begin{align}
\pt \psi_{iI} = \beta^j \partial_j \psi_{iI} + \psi_{jI} \partial_i \beta^j + \alpha \partial_i \Pi_I + \Pi_I \partial_i \alpha.
\end{align}
In the following section, we discuss how we initialize this system of fully general-relativistic equations of motion and how we implement the evolution on a finite grid.

\section{\label{numericalsims} Numerical Simulations}

\subsection{Initial conditions\label{initcond}}

The initial conditions must satisfy the constraint equations, Eqs.~(\ref{Hconstr}) and (\ref{Mconstr}), and be physically motivated. We initialize our parametric resonance scenario in correspondence with the standard semiclassical, linearized treatment of this model's postinflationary dynamics. This choice will facilitate a comparison between {\sc GABERel} and well-known results from CPT, allowing us to analyze the ability of a fully general-relativistic simulation to evolve the nonlinear dynamics and breakdown of coherent oscillations predicted in Ref.  \cite{Easther:2010mr}.

As stated in Sec.~\ref{model}, the mean power in each mode is set by the Bunch Davies vacuum, Eq.~(\ref{bunchdavies}). To realize such initial conditions, we randomly generate Gaussian-distributed power spectra for each component of each field's Fourier mode. The {\sl amplitude} of each Fourier mode is then drawn from a Rayleigh distribution
\beq
 P(\varphi_{Ik}) = \frac{\varphi_{Ik}}{\sigma^2} \rm{e}^{-\varphi_{Ik}^2 /2\sigma^2}, \, 
 \label{rayleigh}
\eeq
where $\sigma^2 = L^3/4\pi^4 \omega_k$. A random realization of Eq.~(\ref{rayleigh}) is then inverse Fourier transformed, giving us an initial field configuration in configuration space.

In order to connect our results with CPT we begin by linearizing the BSSN variables, setting $\gambar = \delta_{ij} + h_{ij}$,  $\Aij = 0 + a_{ij}$, $\phi = 0 +\delta \phi$, $K = K_0 +\delta K$, $ \alpha = 1 + \delta \alpha$, and $\beta^i = 0 + \delta \beta^i$. We set $h_{ij} = 0$, so that gravitational waves do not enter at first order. This choice fixes $\bar{R} = 0$. We also choose $\beta^i =0$ and $a_{ij}=0$. 

To draw a correspondence with Newtonian Gauge perturbation theory, we also have \beq
\label{corresp1}
\delta \alpha = \Phi
\eeq
and 
\beq
\label{corresp2}
\delta \phi = - \Phi/2.
\eeq
Linearizing the Hamiltonian constraint, Eq.~(\ref{Hconstr}), makes it clear that $K_0 = -3H$. Working with the Fourier transform of our variables (denoted by overtildes) this constraint is then equivalent to satisfying
\beq
-k^2 \tilde{\Phi} = \frac{4\pi a^2}{m_{\mathrm{pl}}^2}\delta \tilde{\rho} + a^2 H^2 \delta \tilde{K}
\eeq
which should look identical to the CPT expression, Eq.~(\ref{poisson1}), if we make the identification that 
\beq
\label{defofextcur}
\delta \tilde{K} = 3\left(\dot{\tilde{\Phi}} + H\tilde{\Phi}\right).
\eeq
The choice $a_{ij}=0$ reduces the momentum constraint to 
\beq
\partial_i \delta K =  \frac{12 \pi}{\mplsq}T_{0i}
\eeq
which is equivalent to the CPT expression, Eq.~(\ref{poisson2}), with the same identification, Eq.~(\ref{defofextcur}).

These identifications mean that any solution to the set of perturbation theory equations, Eqs.~(\ref{poisson1}) and (\ref{poisson2}), also satisfy the {\sl linearized} constraints, Eqs.~(\ref{Hconstr}) and (\ref{Mconstr}), so long as we use the identifications, Eqs.~(\ref{corresp1}), (\ref{corresp2}) and (\ref{defofextcur}).  Of course, this is only strictly true at the linear level, but is sufficient for the simulations here, as can be seen in Appendix~\ref{verify}.

\subsection{Other numerical details}

{\sc GABERel} solves differential equations in full numerical relativity on a finite, expanding lattice under periodic boundary conditions. In this work, our lattice resolution is $N^3$ at $N=64$ and we use a time step $\Delta t=\Delta x/20$ (where $\Delta x = L_* / N$ is the initial lattice spacing) with one exception, see Fig.~\ref{rhoslices}. Choosing the time step to be a small fraction of the initial lattice spacing provides good resolution and puts us far from the regime in which causality becomes an issue. 

Simulations on a finite grid eventually reach a stage at which significant power is transferred to the smallest resolvable scales (i.e., the Nyquist frequency, $f_{\rm{Nyq}} = 2\sqrt{3}\pi N / L $). This is dangerous because the difficulty in resolving such modes leads to a gradual increase in numerical inaccuracy, eventually spoiling the validity of the simulation. One way we alleviate this problem is by initializing this system without very much power on these small scales, thereby delaying the onset of this power growth. We do this by smoothly suppressing power in modes with frequencies larger than the cutoff frequency $f_c$, defined by the dimensionless parameter $\xi_c \equiv f_c / f_{\rm{Nyq}} $, which we always choose to be less than 1. This is done by filtering the power in all modes by a window function on the initial slice,
\beq
W(f) = \frac{1}{2}\lp 1- \tanh \left[ \kappa \left( f\frac{L}{2\pi} - \sqrt{3}N \xi_c \right) \right]  \rp
\eeq
Here, $\kappa$ is a parameter between zero and one that determines the sharpness of the window function; in this work we set $\kappa=0.75$ and $\xi_c = 1/8$. In this way we lower power in the smallest resolvable scales, which do not participate in the initial stages of preheating and make it difficult to satisfy the Hamiltonian and momentum constraints. 

\section{\label{Results} Results}

\begin{figure}[ht]
\centering
\includegraphics[width=\columnwidth]{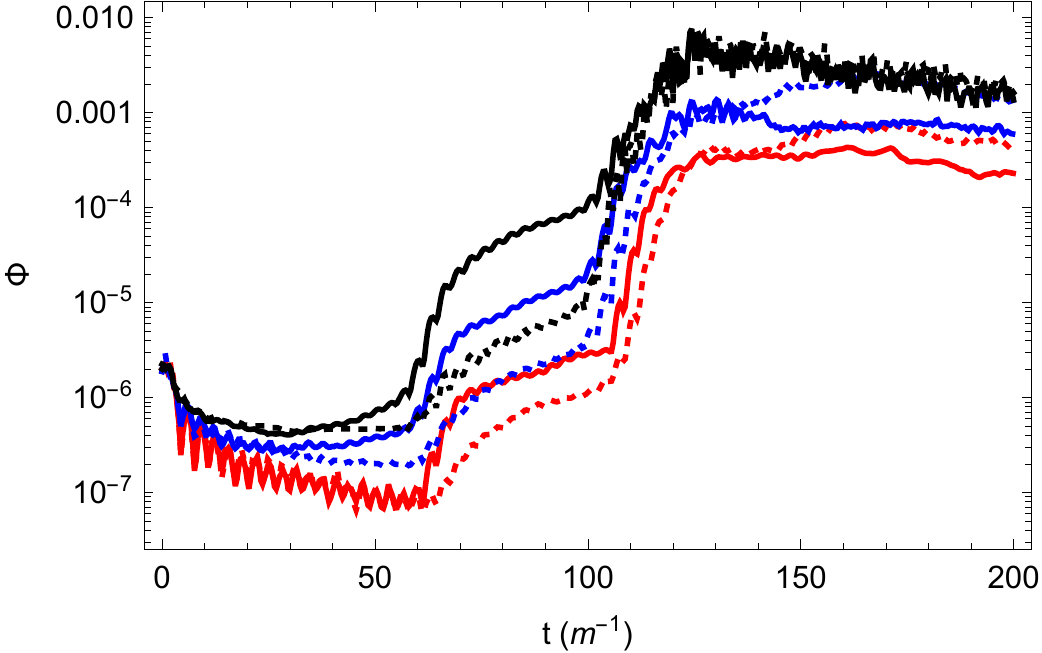} 
\caption{\label{maxmin}The absolute value of the maximum (solid) and minimum (dotted) values of the Newtonian potential $\Phi$ across the box for three different simulations.  The three colors correspond to different box sizes, $L_*=2 \,m^{-1}$ (red), $L_*=5 \,m^{-1}$ (blue), and $L_* = 11 \,m^{-1}$ (black).}
\end{figure}

We begin by examining the effect local gravitational physics has on preheating.  In this case (as in Ref. \cite{Frolov:2008hy}) we expect to see growth of the Newtonian potential during the preheating stages.  We also expect that the statistics of the Newtonian potential will depend on which modes we are able to resolve. As we discussed in Sec.~\ref{CPT}, when the density contrast, $\delta \rho/\left< \rho \right>(k)$ is amplified at the same level, modes near the Hubble scale create larger gravitational deviations from homogeneity than do smaller wavelength disturbances.  Figure~\ref{maxmin} confirms this by showing the maximum and minimum values of the Newtonian potential as a function of time for three different box sizes.  Larger box sizes include longer-wavelength modes which create larger maximum and minimum values of the gravitational potential.  

\begin{figure}[ht]
\centering
\includegraphics[width=\columnwidth]{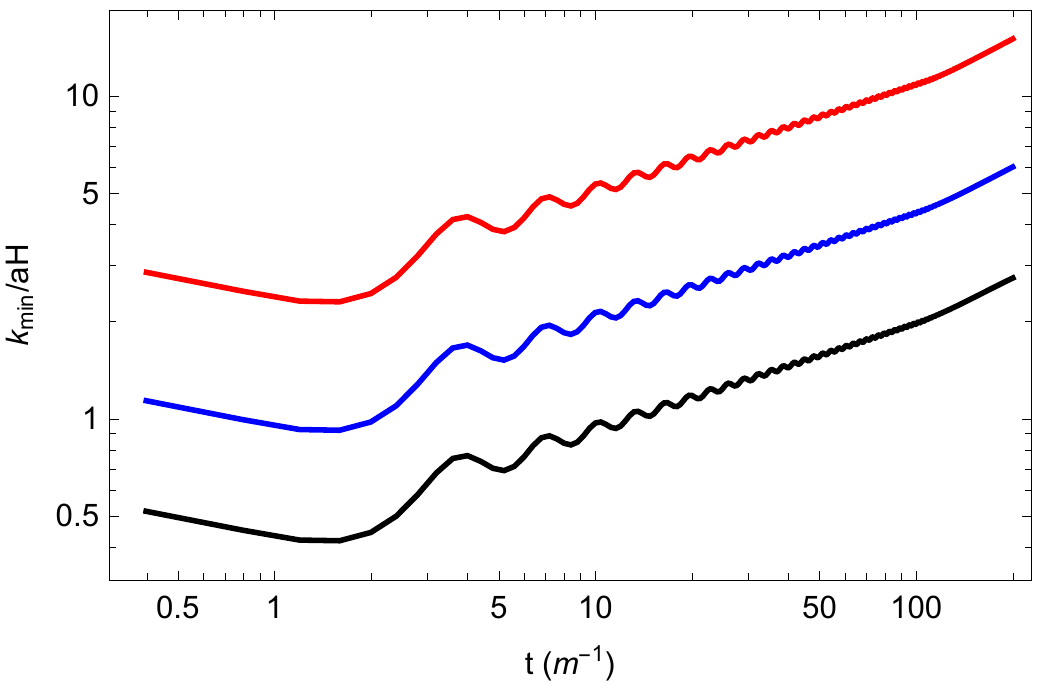} 
\caption{Ratio of the lowest-$k$ mode Newtonian potential to the Hubble rate for three different simulations:  $L_*=2 \,m^{-1}$ (red), $L_*=5 \,m^{-1}$ (blue), and $L_* = 11 \,m^{-1}$ (black).  N.B. the timescale in this figure is logarithmic.}
\label{boxsize}
\end{figure}

The three box sizes we study in Fig.~\ref{maxmin} represent three distinct regimes: the smallest of these sizes, $L_* = 2\,m^{-1}$, is always subhorizon, that is the longest resolvable wavelength mode is bigger than $H^{-1}$ throughout the simulation.  The second case, $L_* = 5\,m^{-1}$, represents a marginal case where the longest wavelength mode becomes superhorizon just at the end of inflation and then quickly retreats inside the Hubble radius.  The third, $L_* = 11\, m^{-1}$, has a long-wavelength mode that stays smaller than $H^{-1}$ for a few oscillations of the scalar field.  We note that the largest of these boxes, $L_*=11\, m^{-1}$ has an initial superhorizon mode which mildly breaks our assumptions of Eq.~(\ref{bunchdavies}). Figure~\ref{boxsize} shows a comparison of the {\sl physical} size ratio of the smallest resolvable wave vector, $k_{\rm min} = 2\pi / a L_*$ compared to the Hubble scale  $H$.   

For the three box sizes featured in Figs.~\ref{maxmin} and \ref{boxsize}, we expect that nonlinear gravity should have a minimal effect on the simulations.  The size of the Newtonian potential does not significantly grow, and we expect to only see small changes in the fields resulting from different realizations of the initial vacuum states.  This is a good regime, then, in which to validate {\sc GABERel} and confirm that no nonlinear gravitational effects enter on these scales.

\begin{figure}[h!]
\centering
\includegraphics[width=.97\columnwidth]{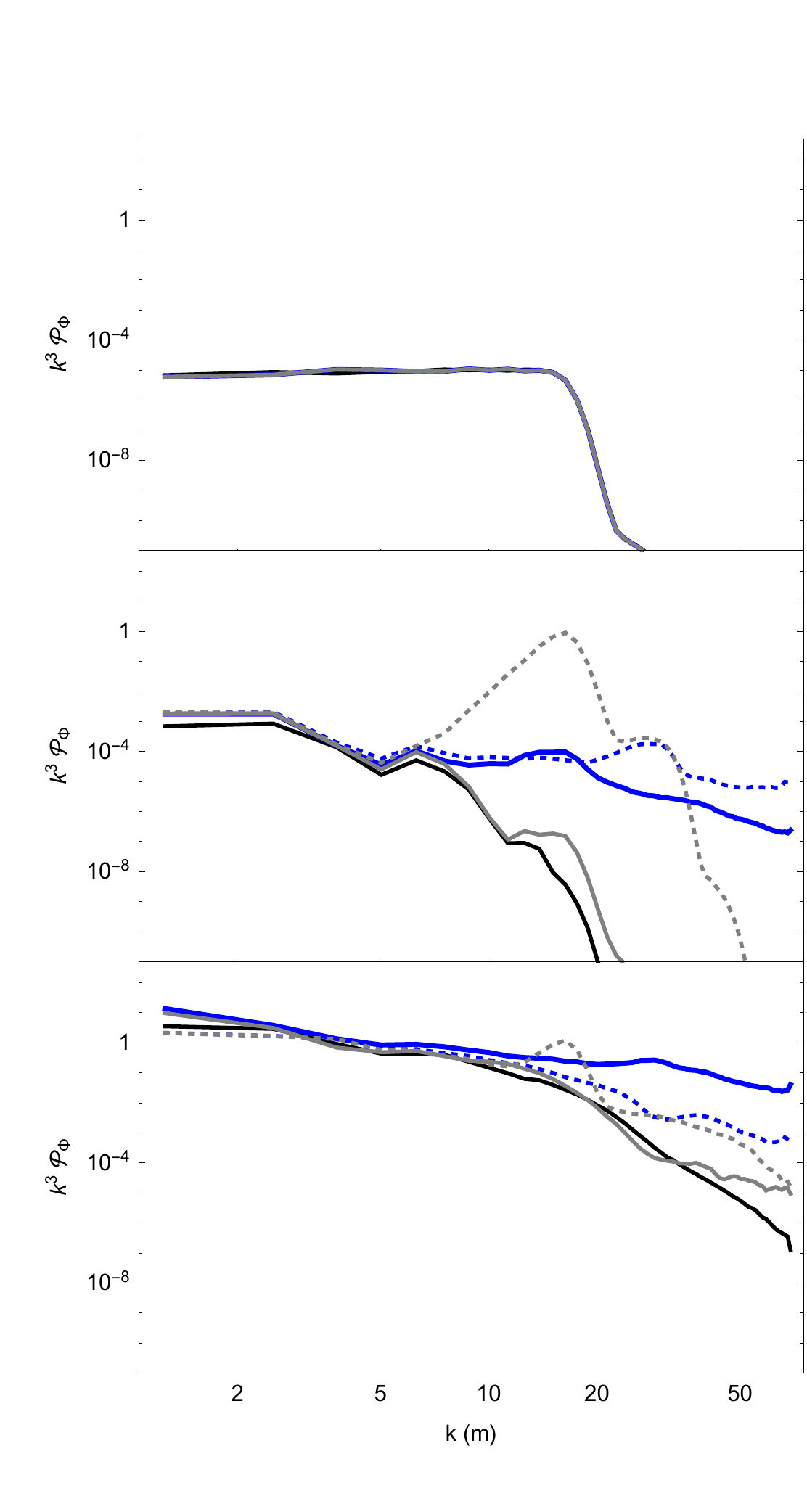} 
\caption{The dimensionless power spectra of the gravitational potentials at 3 different times in 3 different, $L_* =5\, m^{-1}$, simulations: CPT (black), BSSN in $1+\log$ slicing (blue), and BSSN with geodesic slicing (gray). For the BSSN simulations, solid lines correspond to $\Phi_B$ and dashed lines correspond to $\Psi_B$. The top panel is taken at $t=0$ and shows consistency in our initial conditions.  The middle panel, at $t\approx 99 \,m^{-1} $, shows the period of parametric instability, where the spectra of the fields are amplified at certain wavelengths.  The lower panel, at $t\approx 119\, m^{-1}$, shows the spectra at the end of the final stage of preheating, when power is about to move to the Nyquist frequency.   \label{speccompare}}
\end{figure}

To achieve this validation, we look to see if we can calculate the (first-order) gauge-invariant metric perturbations.  Generically, scalar perturbations to the metric can be written as \cite{Weinberg:2008zzc}
\begin{align}
ds^2 =& -(1+2\Phi) dt^2 + 2a(t) B_{,i} dx^i dt  \nonumber \\
&  + a^2(t) \left[(1-2\Psi)\delta_{ij} +2\partial_{ i}\partial_{j}E \right]dx^i dx^j.
\end{align}
In terms of these variables, the Bardeen potentials are (see, e.g., Ref.  \cite{Baumann:2009ds})
\begin{align}
\Phi_B &\equiv \Phi - \frac{d}{dt}\left[a^2\left(\dot{E} - \frac{B}{a}\right)\right]\\
\Psi_B &\equiv \Psi + H a^2\left(\dot{E} - \frac{B}{a}\right)
\end{align}
The translation from the generic form of the metric perturbation to these new variables is straightforward, but computationally expensive and we leave the details on how to calculate these from the BSSN variables to Appendix~\ref{bardeenfromBSSN} (see Ref. \cite{Giblin:2017ezj} for a similar treatment). In CPT, where we have assumed no scalar anisotropic stress, $\Phi = \Phi_B = \Psi_B$, so we would expect the statistics of these gauge-invariant potentials to be identical between our simulations.  To test this, we show a comparison of the power spectra of $\Phi$ (from a CPT simulation) to the dimensionless power spectra, $k^3 \mathcal{P}_\Phi$, as defined by
\beq
\left<0\right|\Phi(\vec{x})\Phi(\vec{y})\left|0\right> = \int \frac{d \ln k}{2\pi^2} k^3\mathcal{P}_\Phi e^{i\vec{k}\cdot(\vec{x}-\vec{y})},
\eeq
of $\Phi_B$ and $\Psi_B$ from a BSSN simulation in Fig.~\ref{speccompare}.  When using a $1+\log$ slicing, we see very good agreement in these quantities in the regions where we anticipate the results to coincide, despite the radically different methods used to calculate these quantities.  The very lowest bins are closest to the horizon where we anticipate CPT and BSSN to disagree, while at higher frequencies nonlinear dynamics mode mix in the BSSN simulations and we see greater power in those simulations.  On small scales, where the power is negligible, differences are due in part to resolving small differences in comparatively large quantities.  We also note that our two situations also differ in that our CPT simulations explicitly ignore scalar anisotropic stress whereas the BSSN ones do not.

To complete our comparison in Fig.~\ref{speccompare}, we also run our code in geodesic slicing.  This slicing condition is known to violate the linearized Einstein equation \cite{Giblin:2018ndw} much more quickly than $1+\log$ slicing, which is confirmed by disagreement in high-frequency modes in Fig.~\ref{speccompare}.  This discrepancy is a sign that the geodesic slicing simulations move out of perturbation theory on some scales faster than those in $1+\log$ slicing.  Finally, we note that we have to take extreme care when analyzing the output of the BSSN simulations (see Appendix~\ref{verify}).  In this code, when power is moved to the Nyquist frequency, we see that the Hamiltonian and momentum constraints, Eqs.~(\ref{Hconstr}) and (\ref{Mconstr}), are no longer satisfied.  We take care not to derive physical meaning from these simulations after this time.  For all of the BSSN simulations we present here, the constraints start to grow around $t=110\,m^{-1}$ and become unsatisfactorily satisfied before $t=130\,m^{-1}$.

\begin{figure*}[h!]
\centering
\includegraphics[width=\textwidth]{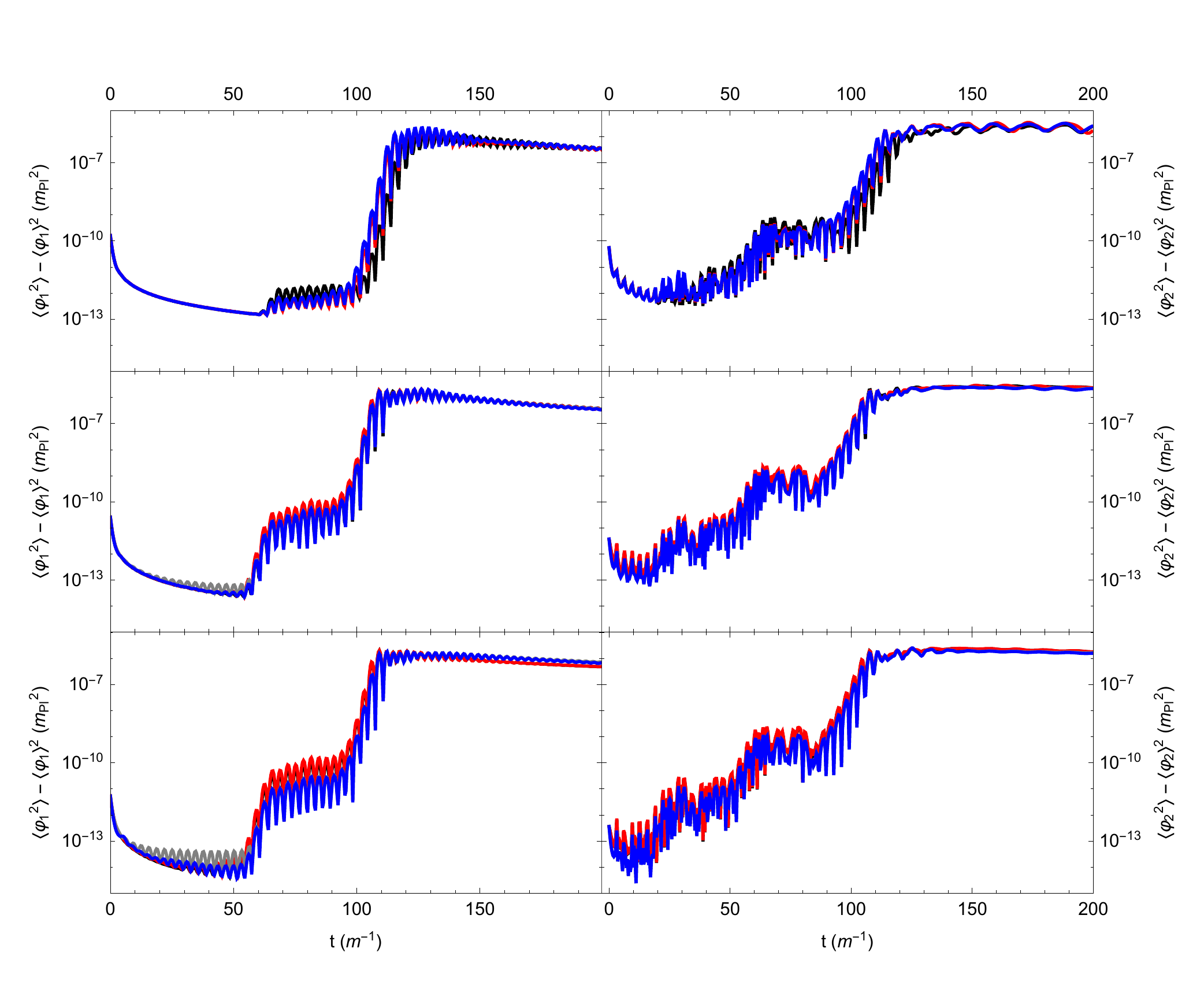} 
\caption{Comparison of variances of the inflaton $\varphi_1$ and matter field $\varphi_2$ as a function of time for three different simulations:  FLRW (black), CPT (red), BSSN in $1+\log$ slicing (blue) and BSSN in geodesic slicing (gray).  N.B. simulations for $L_*=2\,m^{-1}$ do not include geodesic slicing runs. Simulations are at box sizes of $L_* = 2\,m^{-1}$ (top panels), $L_* = 5\,m^{-1}$ (middle panels), and $L_* = 11\,m^{-1}$ (bottom panels).  \label{6panel}}
\end{figure*}

Additionally, one can look at the statistics of the two fields to see agreement between our simulations.  Figure~\ref{6panel} shows a comparison of the variances of the two scalar fields---and confirms that there are no notable differences between the simulations we run across box sizes and methods.  At some box sizes, geodesic slicing alone exhibits early oscillations in the inflaton variance during the phase of coherent oscillations. This demonstrates the importance of slicing conditions when comparing fully relativistic results to those from CPT and FLRW treatments.

\begin{figure}[ht]
\centering
\includegraphics[width=\columnwidth]{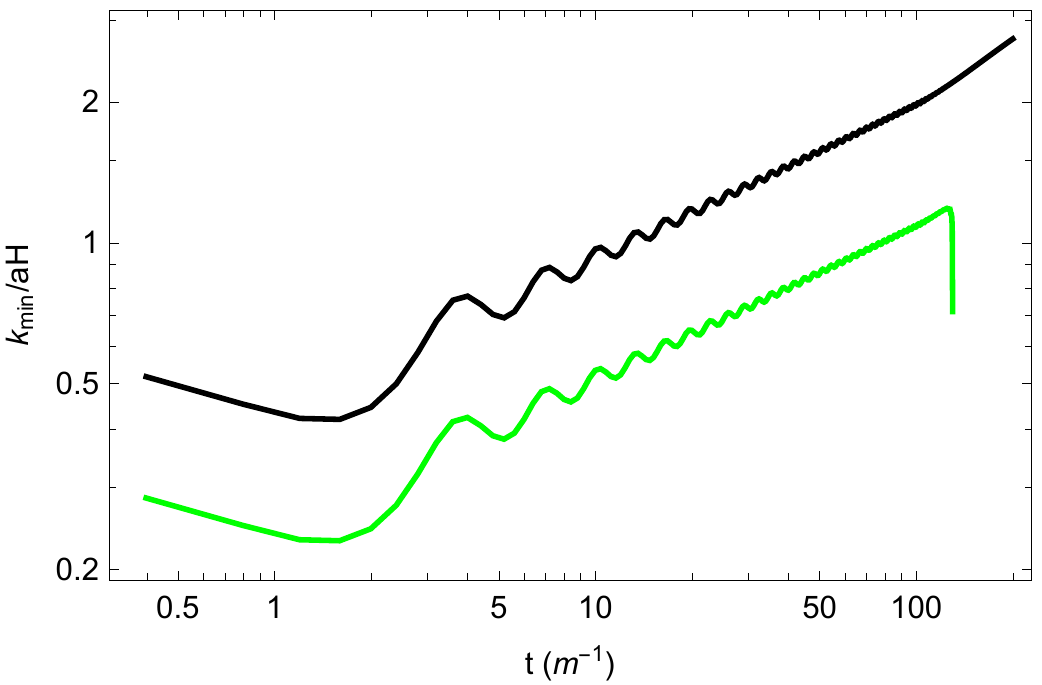} 
\caption{Ratio of the lowest-$k$ mode Newtonian potential to the Hubble rate for two CPT simulations: $L_* = 11 \,m^{-1}$ (black) and $L_* = 20 \,m^{-1}$ (green). The simulation with the largest box (and hence the largest wavelength mode) crashes shortly after $\Phi_{k_{\rm{min}}}$ enters the horizon.  N.B. the timescale in this figure is logarithmic.}
\label{boxsizev2}
\end{figure}

\begin{figure}[ht]
\centering
\includegraphics[width=\columnwidth]{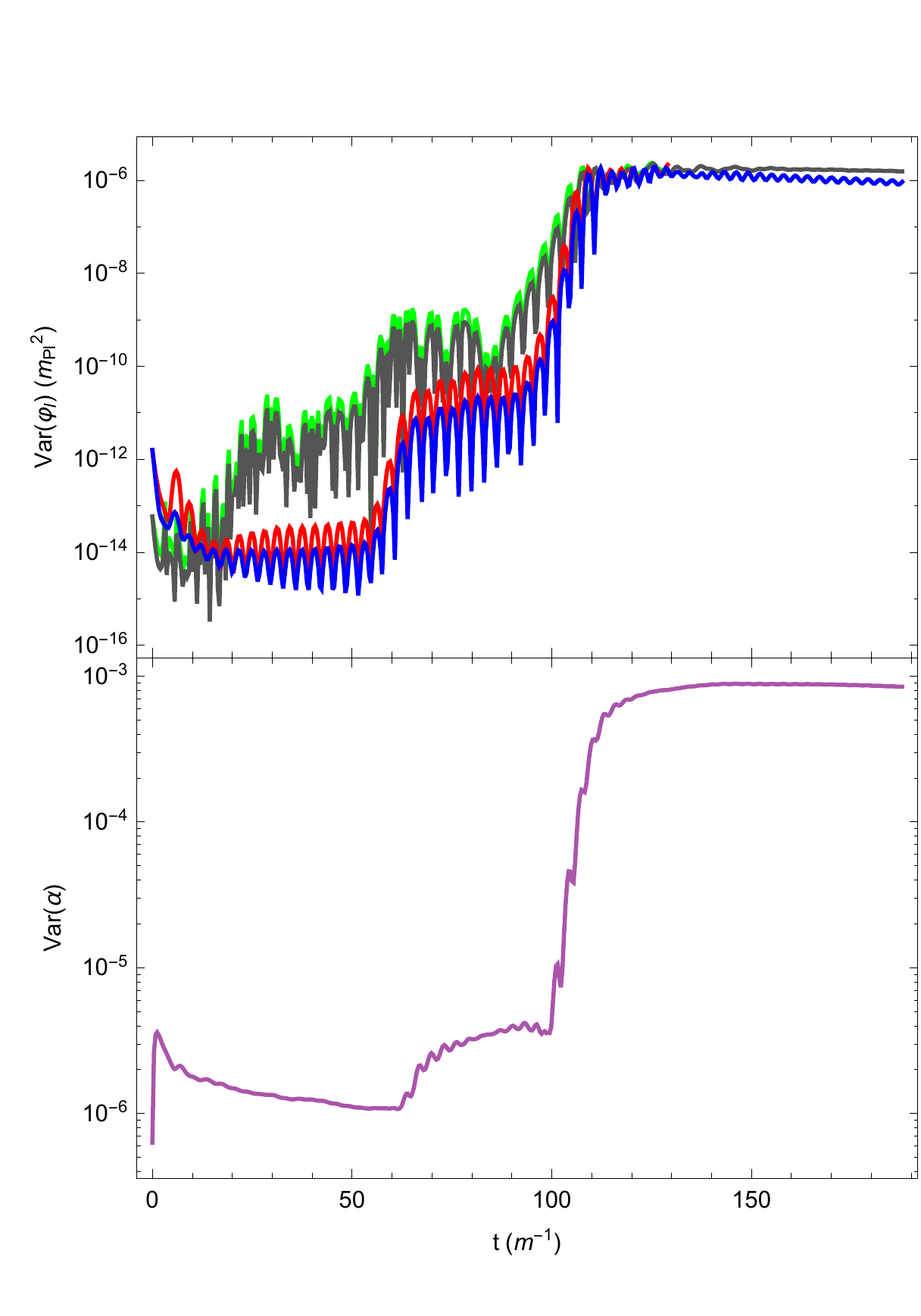}
\caption{\label{variance20}The top panel shows a comparison of variances of the inflaton $\varphi_1$ and matter field $\varphi_2$ as a function of time for three different simulations:  CPT ($\varphi_1$ is red, $\varphi_2$ is green), and BSSN in $1+\log$ slicing ($\varphi_1$ is blue, $\varphi_2$ is gray). The bottom panel shows the variance of $\alpha$ for the BSSN simulation; $\alpha$ has a homogenoues value that stays at $\langle\alpha\rangle = 1\pm 10^{-5}$ throughout the simulation.}
\end{figure}

We can now move on to see if there are indications of stronger gravitational instabilities if we include more horizon-sized modes for a longer time in the simulation.  For this we consider a box size of $L_* = 20$, so that we have more modes near the Hubble scale for a longer portion of the simulation. We note that our initial conditions are also slightly inconsistent in this scenario, i.e., a small number of these modes that are not well within the horizon and are not exactly Bunch-Davies, Eq.~(\ref{bunchdavies}). Figure~\ref{boxsizev2} shows the ratio of the physical size of the lowest-frequency mode to the Hubble scale for a box of size $L_*=20$ along with a comparison to the same quantity for $L_* = 11$.  
We see in this case that the lowest frequency mode stays outside the horizon for more of the simulation, allowing many modes to cross into the horizon during the preheating process.  These modes do become inhomogenous as the simulation proceeds and, as they are entering the horizon, cause the Newtonian potential to grow to order unity and the simulation to crash.  As predicted in Ref. \cite{Easther:2010mr}, these additional modes become nonlinear during the phase of coherent oscillations; the Universe may undergo a number of $e$-foldings before the contribution to the extrinsic curvature from the shift becomes comparable to that from the homogeneous evolution.  This amplification is exactly the effect we see in our CPT simulations.\footnote{The authors of Ref. \cite{Easther:2010mr} work in spatially flat gauge where the shift contributes to the Bardeen potentials. We then rely on comparing the (first-order) gauge invariant potentials either $\Phi$ (from our CPT simulations) or $\Phi_B$ and $\Psi_B$---to analyze this predicted breakdown of linearity.}
BSSN simulations, on the other hand, do not rely on the assumptions of CPT, and, hence, the code does not crash at this point.  Figure~\ref{variance20} shows a comparison between the CPT simulation and the BSSN simulations for $L_* = 20\,m^{-1}$ which demonstrates that fully nonlinear gravity is able to resolve the moderate density contrasts that exist as these modes are entering the horizon.  It is important to note, though, that when we move to full GR, we do not see a breakdown of the coherent oscillations of the inflation field, despite the fact that the Newtonian potential (or the Bardeen potentials) grows to be close to order unity.

The bottom panel of Fig.~\ref{variance20} shows that statistics of the lapse, $\alpha$ for this simulation.  This quantity is important when probing the existence of strong gravity and the possible creation of primordial black holes.  In the simulations we have run here we see minimal departure in the lapse, and hence, do not see the creation (or seeds) of primordial black holes.  This statement is, however, a model-dependent statement and not a generic prediction for all models of preheating.

\begin{figure*}[ht]
\centering
\includegraphics[width=\textwidth]{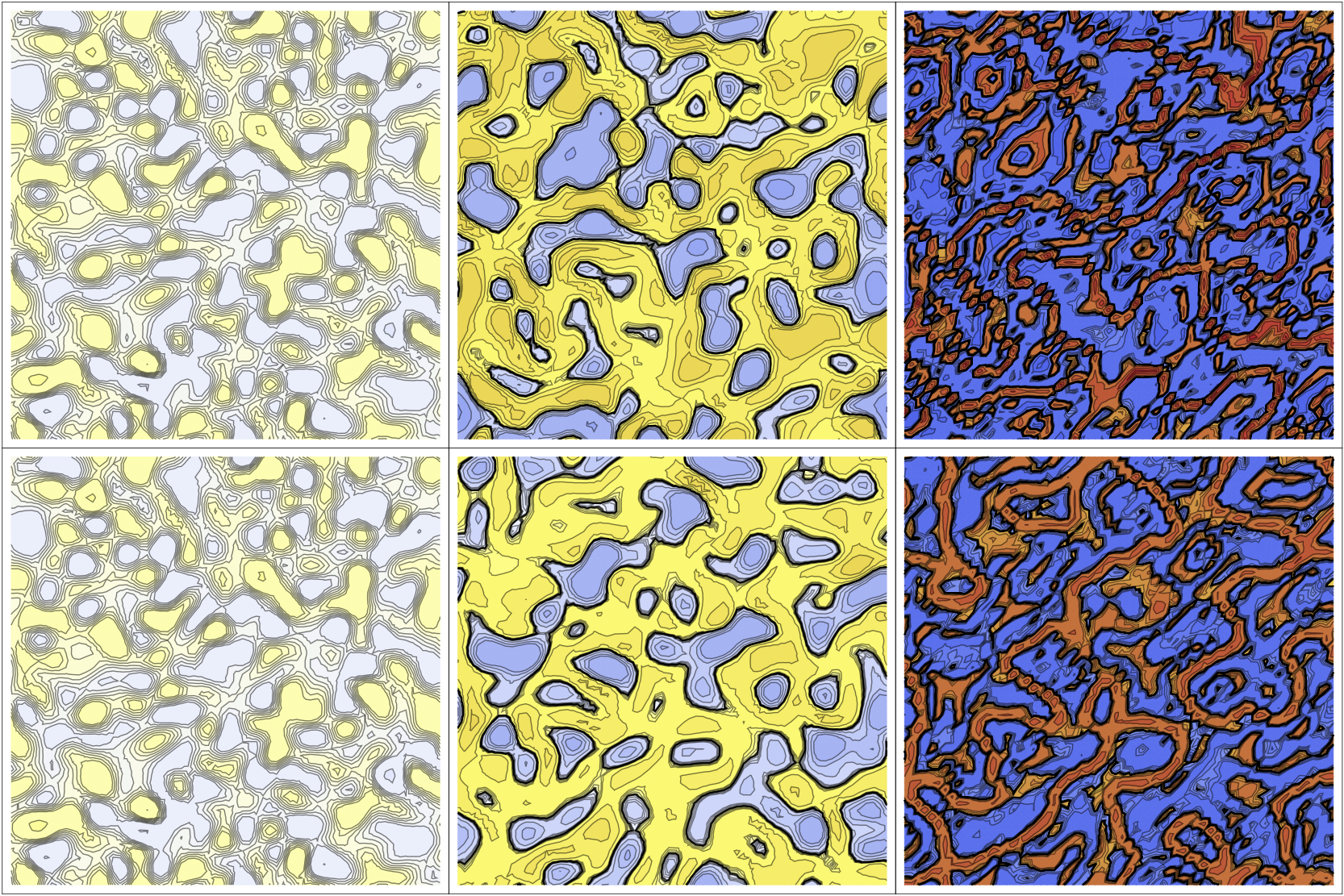}
\caption{\label{rhoslices}Chronological two-dimensional slices of the density contrast $\delta=\delta \rho / \left< \rho \right>$ in $L_* = 20\,m^{-1}$ simulations in CPT (top panels) and {\sc GABERel} (bottom panels). From left to right the slices are at $t=0 $, $t\approx 87\,m^{-1}$, and $t\approx 121\,m^{-1}$. For this final time, the maximum (minimum) values of the density contrast are $\delta \rho/\left<\rho \right> \approx 1.43$  ($-0.76$) for the CPT simulation and $\delta \rho/\left<\rho\right> \approx 0.73$ ($ -0.40$) for the BSSN simulation. Here we use $\Delta t=\Delta x/80$; this smaller time step ensures that, for the larger box sizes used in this figure, the simulation runs until the dynamics become truly nonlinear in the final panel.}
\end{figure*}

We can also look towards the density contrast, $\delta=\delta \rho / \left< \rho \right>$, to assess the onset of nonlinear gravitational physics. This provides an intuitive way of assessing the validity of linear perturbation theory and a consistency check with Fig~\ref{boxsizev2}. CPT is built on the assumption that $\delta$ is small. In Fig.~\ref{rhoslices} we compare two-dimensional slices of $\delta$ throughout a CPT and a BSSN simulation for $L_* = 20\,m^{-1}$. For both simulations, the density contrast initially grows slowly and remains small, and CPT remains valid.  However, around $t \approx 120\,m^{-1}$---right after $\Phi_{k_{\rm{min}}}$ enters the horizon---$\delta$ grows sharply to order 1 in the CPT and BSSN simulations, with the growth being only marginally more pronounced in the CPT simulation. The CPT simulation fails at this point. This confirms the breakdown of linearity predicted in Refs. \cite{Bassett:1998wg,Bassett:1999mt,Bassett:1999ta,Tsujikawa:2002nf} (and reinforced in Ref. \cite{Easther:2010mr}) and shown in Fig.~\ref{boxsizev2}: once the horizon-sized modes enter the horizon during preheating, the density contrast quickly exceeds unity and the linear treatment is no longer valid, necessitating a fully general-relativistic treatment. Despite this nonlinear instability, we do not observe strong gravitational collapse or the formation of PBH during this stage of preheating. The growth of the density contrast moves out of the regime of validity of (first-order) CPT after this point, but does not necessarily signal unbounded growth of overdensities. However, neither the CPT nor BSSN simulation is valid shortly after this point. The former because $\Phi$ becomes large and the simulation crashes, the latter because power is moved to the Nyquist frequency and the constraints are no longer satisfied. Still, there is no indication that the gravitational instabilities are large enough, in this model, to continue to collapse. Nevertheless, thanks to the robustness of full numerical relativity, {\sc GABERel} is able to successfully resolve these nonlinear physical processes that are inaccessible to CPT.

\section{Discussion \label{discussion}}

We have explored the effect of adding local gravitational effects to a toy model of preheating and present a rigorous analysis of gravitational effects after inflation.  For this model of preheating, we have shown that there are no unexpected results that arise when local gravity---in either perturbation theory or in full GR---is included in the dynamics of the Universe.  When local gravity is included, we see the excitation of scalar perturbations; in these cases, the Newtonian potential can get as large as $\Phi \sim 10^{-1}$, however, the subsequent field dynamics do not allow for these perturbations to become nonlinear and there is no indication of compact structures forming.  While the Newtonian potential does grow, it does not seem that this is enough to cause nonlinearity in the gravitational part of the system---even when the Newtonian potential becomes order unity.  The variations of the Newtonian potential are on large scales and do not seem to cause gravitational collapse.

We have also introduced a new computational tool, {\sc GABERel}: software that can evolve scalar fields in full numerical relativity.  We have validated that this software reproduces expectations from perturbation theory in its regime of validity and can be extended to regions in which perturbation theory fails.  Using this tool for a canonical model of preheating shows that nonlinear gravitational effects stabilize the instabilities present in perturbation theory.  We see that the phenomena of preheating in this setup shares a background evolution with the unperturbed FLRW analysis for the entirety of the simulation, even when departures grow beyond first-order perturbation theory.  It also shows that some choices of coordinates (slicing) do a better job satisfying the linearized Einstein's equations \cite{Weinberg:2008zzc,Giblin:2018ndw} than do others.  

Finally, it is important to note that the lack of the formation of collapsed structures is not a generic result. In addition to previous CPT results hinting towards linear gravitational physics, the model of preheating studied here, Eq.~(\ref{preheatmodel}), is not known to cause large metric perturbations.  We envision that stronger instabilities (such as superhorizon or tachyonic instabilities \cite{Bassett:1999cg,Bassett:2000ha,Finelli:2000ya,Felder:2000hj,Felder:2001kt,GarciaBellido:2001wn,Kofman:2001rb,GarciaBellido:2001cb,Copeland:2002ku,Suyama:2004mz,Suyama:2006sr,Barnaby:2006cq,Felder:2006cc,DiazGil:2007qx,DiazGil:2008tf,Dufaux:2008dn,Battefeld:2009xw,Barnaby:2009wr,Abolhasani:2009nb,Braden:2010wd,Antusch:2015vna,Adshead:2016iae,DeCross:2016cbs,Tranberg:2017lrx,Adshead:2017xll}) that are closer to the horizon would be ideal candidates for more dramatic gravitational physics, and we look forward to studying more of these models in the future.

\section{Acknowledgements}

We are extremely grateful for a number of enlightening conversations as this work was being completed.  We are indebted to Thomas Baumgarte and James Mertens for discussion of the BSSN formalism and helping with code validation techniques.  We are also grateful for many conversations with Peter Adshead, Leia Barrowes, Robert R. Caldwell, David I. Kaiser, and Zachary J. Weiner regarding cosmological perturbation theory and for comments on our drafts.  We also acknowledge the undergraduates responsible for probing cosmological perturbation theory in different contexts that eventually yielded the computational expertise to make this work possible: Arthur Conover, Maggie Murphree, Rachel Nguyen, Caroline Popiel, and Christian Solorio.  J.T.G. is supported by the National Science Foundation Grant No. PHY-1719652.  Numerical simulations here were performed on hardware purchased by the National Science Foundation, Kenyon College, and the Kenyon College department of physics, and we would like to acknowledge the infrastructure support provided by Kenyon College.

\appendix

\section{Bardeen variables}
\label{bardeenfromBSSN}

In Sec.~\ref{Results} we relied on the ability to write the Bardeen potentials strictly in terms of BSSN variables. To calculate these we draw a correspondence between the BSSN metric, Eq.~(\ref{BSSNmet}), and a general form of a metric with scalar perturbations  \cite{Weinberg:2008zzc}
\begin{align}
ds^2 =& -(1+2\Phi) dt^2 + 2a(t) B_{,i} dx^i dt \nonumber \\
&  + a^2(t) \left[(1-2\Psi)\delta_{ij} +2\partial_{ i}\partial_{j}E \right]dx^i dx^j
\end{align}
In terms of these variables, the Bardeen potentials are \cite{Baumann:2009ds}
\begin{align}
\Phi_B &\equiv \Phi - \frac{d}{dt}\left[a^2\left(\dot{E} - \frac{B}{a}\right)\right] \label{bardpotphi}\\
\Psi_B &\equiv \Psi + H a^2\left(\dot{E} - \frac{B}{a}\right)\label{bardpotpsi}
\end{align}
We first set the purely spatial parts of the metric to be equal in both gauges, $g_{ij}^{\rm{BSSN}}=g_{ij}^{\rm{CPT}}$
\begin{equation}
\label{identifyhere}
 a^2(t) \left[(1-2\Psi)\delta_{ij} +2\partial_i\partial_jE \right] = e^{4\phi} \bar{\gamma}_{ij}
\end{equation}

There are two relationships that we can get from this. First, by taking the trace of both sides, where repeated lower indices implies a sum, we get
\begin{equation}
a^2[3-6\Psi+2\partial_i \partial_i E] = \gamma_{ii}.
\end{equation}
We can solve this for $\nabla^2E$,
\begin{equation}
    \nabla^2 E =\left[3\Psi - \frac{3}{2} + \frac{\gamma_{ii}}{2 a^2}\right]
    \label{nablaE}
\end{equation}
or in Fourier space variables (denoted by overtildes)
\begin{equation}
    \tilde{E} = -\frac{1}{k^2}\left[3\tilde{\Psi} + \frac{\tilde{\gamma}_{ii}}{2 a^2}\right]
\end{equation}
where we have dropped the $3/2$ since it only contributes to the homogeneous mode which must be zero for a perturbation. We can also take the mixed second derivative of Eq.~(\ref{identifyhere}), $\partial_i \partial_j$, where again we sum over repeated downstairs indices,
\begin{equation}
    a^2\left[-2\nabla^2 \Psi + 2 \nabla^2\nabla^2 E \right] = \partial_i\partial_j \gamma_{ij}
\end{equation}
which can be rearranged
\begin{equation}
    \nabla^2\left[\nabla^2 E -  \Psi\right] = \frac{\partial_i\partial_j \gamma_{ij}}{2a^2}
\end{equation}
and simplified by using Eq.~(\ref{nablaE}),
\begin{equation}
    \nabla^2\left[\left[3\Psi + \frac{\gamma_{ii}}{2 a^2}\right] -  \Psi\right] = \frac{\partial_i\partial_j \gamma_{ij}}{2a^2}
\end{equation}
\begin{equation}
    \nabla^2\left[2\Psi +  \frac{\gamma_{ii}}{2 a^2}\right] = \frac{\partial_i\partial_j \gamma_{ij}}{2a^2}
\end{equation}
This yields an expression for $\nabla^2 \Psi$,
\begin{equation}
    \nabla^2 \Psi = \frac{1}{4}\left[\frac{\partial_i\partial_j \gamma_{ij}}{a^2} - \frac{\nabla^2 \gamma_{ii}}{ a^2}\right].
\end{equation}
or in Fourier space,
\begin{equation}
 \tilde{\Psi} = \frac{1}{4}\left[\frac{k_ik_j}{k^2}\frac{\tilde{\gamma}_{ij}}{a^2} - \frac{\tilde{\gamma}_{ii}}{a^2}\right].
\end{equation}
This allows us to calculate $E$ strictly in terms of BSSN variables and $a$ (which will later be replaced by $\left< \rm{e}^{4 \phi} \right>$),
\begin{equation}
    \tilde{E} = -\frac{1}{k^2}\left[\frac{3}{4}\frac{k_ik_j}{k^2}\frac{\tilde{\gamma}_{ij}}{a^2}  - \frac{1}{4}\frac{\tilde{\gamma}_{ii}}{a^2}\right]
\end{equation}
The last thing we need is $B$, which we get from $g_{0i}$,
\begin{equation}
2a\partial_i B = 2 \beta_i 
\end{equation}
Introduce a sum by differentiating both sides,
\begin{equation}
a\partial_i \partial_i B = \partial_i \beta_i 
\end{equation}
which in Fourier space becomes
\begin{equation}
\tilde{B} = -\frac{1}{a} \frac{i k_j \tilde{\beta}_j}{k^2}.
\end{equation}

The scalar metric perturbations in terms of $a$ and BSSN variables can be summarized
\begin{align}
    \tilde{E} &= -\frac{1}{k^2}\left[\frac{3}{4}\frac{k_ik_j}{k^2}\frac{\tilde{\gamma}_{ij}}{a^2}  - \frac{1}{4}\frac{\tilde{\gamma}_{ii}}{a^2}\right]\\
\tilde{B} &= -\frac{1}{a} \frac{i k_j \tilde{\beta}_j}{k^2} \\
\Phi &= \alpha - 1 \\
\tilde{\Psi} &= \frac{1}{4}\left[\frac{k_ik_j}{k^2}\frac{\tilde{\gamma}_{ij}}{a^2} - \frac{\tilde{\gamma}_{ii}}{a^2}\right].
\end{align}
where repeated indices are summed over. We define the term common in both Bardeen potentials, Eqs.~(\ref{bardpotphi}) and  (\ref{bardpotpsi}), as $\mathcal{B} \equiv a^2\left(\dot{E} - B / a \right)$. In BSSN variables we then have
\begin{align}
    \mathcal{B} &= \frac{3}{2}\frac{k_ik_j}{k^4}H \gamma_{ij} - \frac{3}{4}\frac{k_ik_j}{k^4} \dot{\gamma}_{ij}
    - \frac{1}{2}\frac{1}{k^2}H \gamma_{ii}\nonumber \\
    & \, \, \, \, \, \, \, + \frac{1}{4}\frac{1}{k^2} \dot{\gamma}_{ii}
    +  \frac{i k_i \beta_i}{k^2},
\end{align}
where we the Hubble rate with the average extrinsic curvature $H = -\left<K\right>/3$ as we did when linearizing the Hamiltonian constraint, Eq.~(\ref{Hconstr}). The Bardeen potentials are then computed strictly in terms of BSSN variables,
\begin{align}
\Phi_B &\equiv \alpha -1  - \dot{\mcb} \label{bardphibssn}\\
\Psi_B &\equiv \frac{1}{4}\left[\frac{k_ik_j}{k^2}\frac{\gamma_{ij}}{a^2} - \frac{\gamma_{ii}}{a^2}\right] - \frac{\left< K \right>}{3} \mcb \label{bardpsibssn}.
\end{align}
The time derivative in Eq.~(\ref{bardphibssn}) is very cumbersome, and we do not need to evolve the Bardeen potentials as they are only used to compare our BSSN results with CPT computations. Therefore it is sufficient to store $\mcb$ from the previous step and calculate $\dot{\mcb}$ with a Euler method. Incidentally, this means we cannot calculate  $\dot{\mcb}$ until the second time step and also means we introduce a fixed numerical error of order $dx^2$ to $\Phi_B$.

\section{Code Verification}
\label{verify}

\begin{figure*}[h!]
\centering
\includegraphics[width=.8\textwidth]{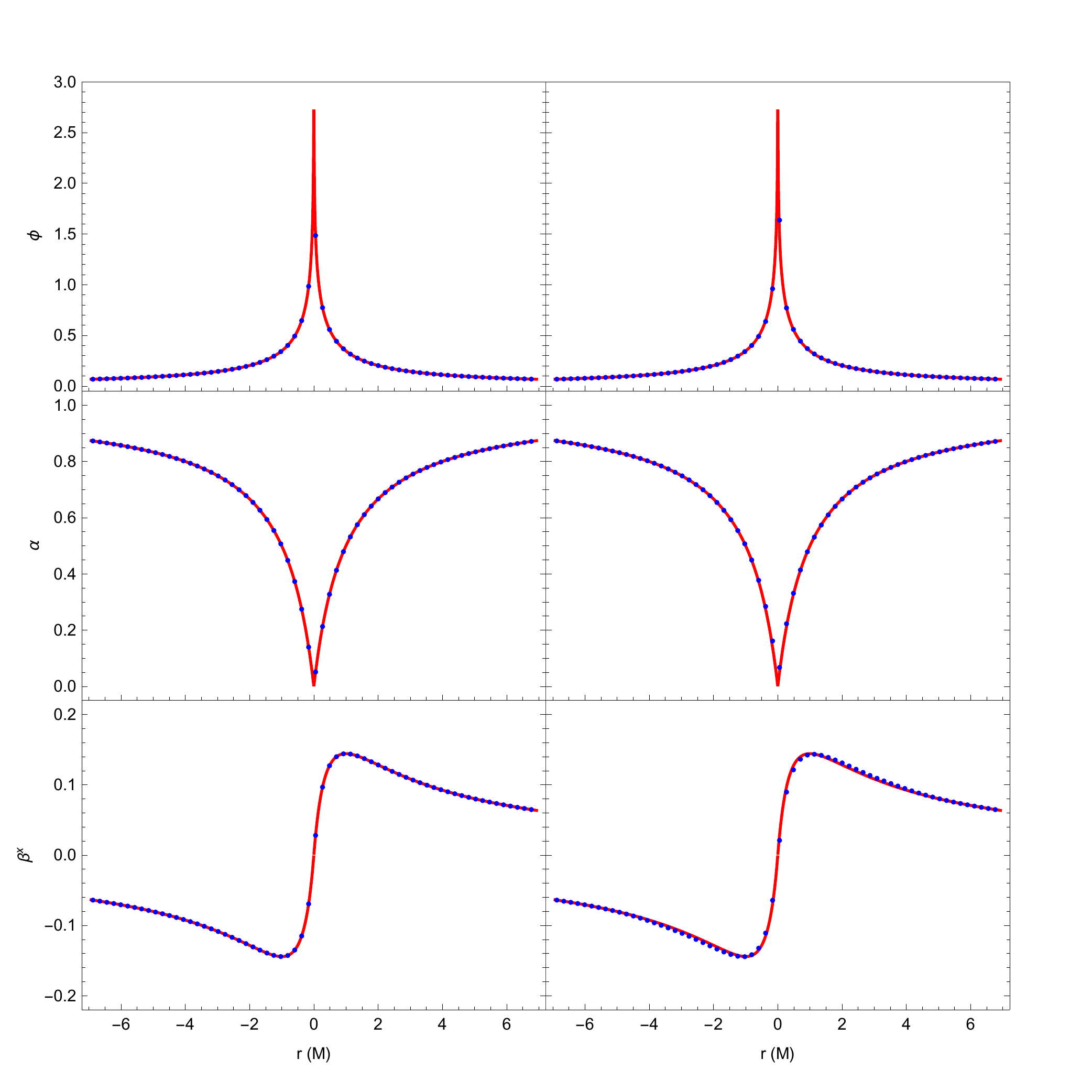} 
\caption{The conformal factor $\phi$, lapse $\alpha$, and radial component of the shift $\beta^r$, for slices of the Schwarzschild black hole described in Appendix~\ref{verify}.  Note that the one-dimensional slice is taken thorough the diagonal of the box, so that it can cut through the center of the black hole---which is placed between grid points near the middle of the box.  The left three panels are at $t=0$ and the right three panels are at $t=10 \, M$. \label{BHfig}}
\end{figure*}

We have run a number of tests on {\sc GABERel} to verify \cite{Szilagyi:1999nu} that our gravitational evolution is accurate.  Among these are the robust stability test and a Schwarzchild black hole test. In the former we find no exponential noise growth and the constraint violations remain bounded for long-term evolution. For the latter, we simulated a Schwarzschild geometry in {\sl trumpet} coordinates \cite{Dennison:2014sma}.  In this setup, the conformal factor $\phi$, lapse $\alpha$ and shift $\beta^i$ take the form,
\begin{align}
\psi \equiv& \log \phi =  \sqrt{1+\frac{1}{r}}, \\ 
\alpha =& \frac{r}{1+r}, \\
\vec{\beta} =& \frac{r}{\left(r+1\right)^2}\hat{r} ,
\end{align}
which corresponds to a choice of $R_0 = M$; N.B. for this test, we use units where $G=1$ and, hence, space and time are in units of $M$. We chose this particular slicing because all of the degrees of freedom of the extrinsic curvature, $K$ and $\tilde{A}_{ij}$, are nonzero throughout the simulation, even though the solution is static. The black hole is initially centered $dx/4$ from a central lattice point. This intentional asymmetry constitutes a more general and robust test of our code's gravitational dynamics than would otherwise be the case if we impose some artificial symmetry on the system. We evolve the gauge variables, $\alpha$ and $\beta^i$, according to $1+\log$ and $\eta=0$ Gamma-driver conditions, respectively,  with advective shift terms,
\begin{align}
(\partial_t -\beta^j\partial_j) \alpha &= -\alpha\left(1-\alpha\right)K \\
(\partial_t -\beta^j\partial_j) \beta^i &= \frac{3}{4}\bar{\Gamma}^i .
\end{align}
While the rest of our simulation uses central finite differencing in an RK4 scheme, anything evolved with advective shift terms uses {\sl upwind derivative stencils}.  Gamma driver serves to approximate the more difficult to evolve Gamma-freezing condition, $\partial_t \christoff = 0$.  Figure~\ref{BHfig} shows the stability of our code to this setup. Given our simulation's periodic boundary conditions and the asymmetric initial position of the singularity, we do not necessarily expect that our solution should approach a perfectly steady state after long-term evolution. Nevertheless we observe no noticeable changes between the initial slice and $t=10\,M$. In this simulation, the box is taken to be $L=8\,M$ with $N=64^3$ points.

\begin{figure}[ht]
\centering
\includegraphics[width=\columnwidth]{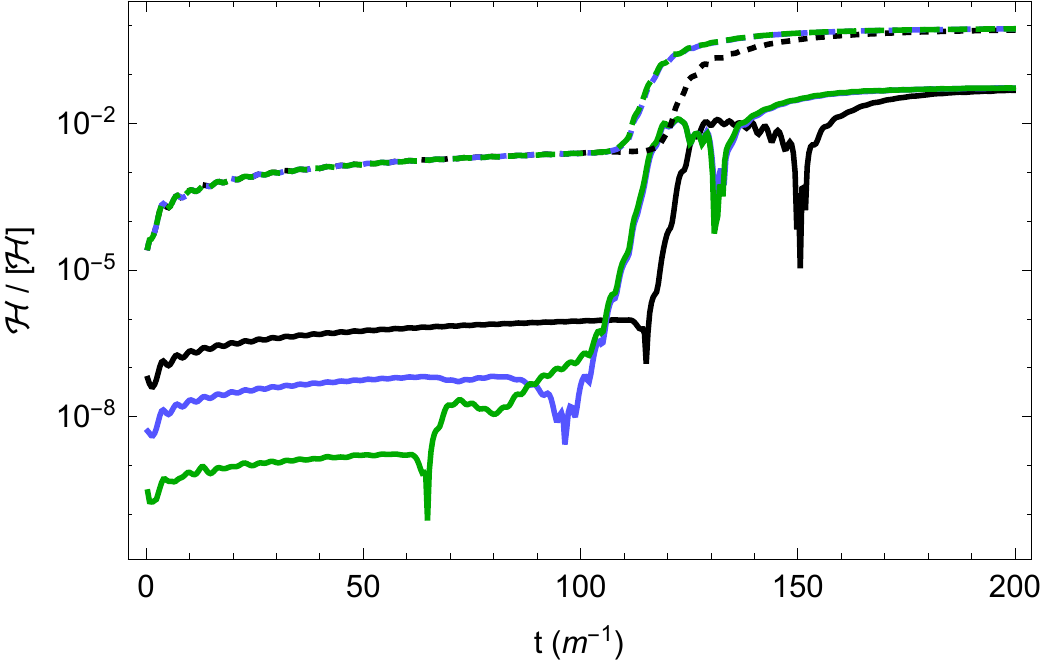} 
\caption{\label{dtconverge}Testing numerical convergence via the average, $\langle\mathcal{H}\rangle/\left[\mathcal{H}\right]$ (solid), and rms, $\sqrt{{\rm Var}(\mathcal{H}/\left[\mathcal{H}\right])}$ (dashed), Hamiltonian constraint violation in three different $L_* = 11 \,m^{-1}$ simulations: $dt=dx/10$ (black), $dt=dx/20$ (blue), $dt=dx/40$ (green).}
\end{figure}

\begin{figure}[ht]
\centering
\includegraphics[width=\columnwidth]{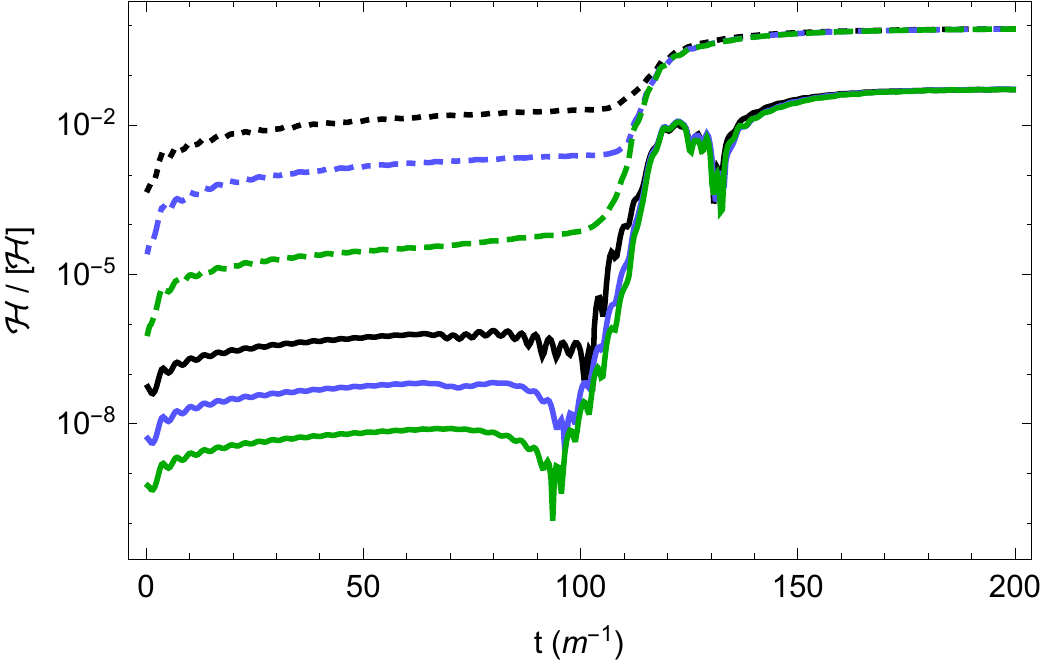} 
\caption{\label{cutoffconverge} Testing numerical convergence via the average, $\langle\mathcal{H}\rangle/\left[\mathcal{H}\right]$ (solid), and rms, $\sqrt{{\rm Var}(\mathcal{H}/\left[\mathcal{H}\right])}$ (dashed), Hamiltonian constraint violation in three different $L_* = 11 \,m^{-1}$ simulations: $\xi_c = 1/4$ (black), $\xi_c = 1/8$ (blue), $\xi_c = 1/16$ (green).}
\end{figure}

It's also important for us to verify that we can trust the code for the simulations presented in Sec.~\ref{Results}.  The constraints \ref{Hconstr} and \ref{Mconstr} should remain small and bounded throughout the system's evolution. However, the constraints are dimensionful, and so we must normalize them.  Here we will show the Hamiltonian constraint and normalize it by the root sum of the squares of the terms in the constraint, 
\begin{align}
\left[\mathcal{H}\right]  \equiv& \Biggl[ \lp \bar{\gamma}^{ij} \bar{D}_i \bar{D}_j e^{\phi} \rp ^2 + \lp \frac{e^{\phi}}{8}\bar{R} \rp ^2 +  \lp \frac{e^{5\phi}}{8}\Aiju \Aij \rp ^2 \nonumber\\ 
& + \lp  \frac{e^{5\phi}}{12}K^2 \rp ^2 + \lp  2\pi e^{5\phi} \rp ^2  + \lp 2\pi e^{5\phi} \rho \rp ^2 \Biggr] ^{1/2}.
\end{align}
We show the evolution of the Hamiltonian constraint in Figs.~\ref{dtconverge} and \ref{cutoffconverge}. It is not surprising that the constraint grows in the third phase of preheating as this is precisely when significant power is transferred to the smallest resolvable scales and numerical error grows. After this point, the numerical evolution is stable but no longer reliable. 


\end{document}